\documentclass[pra,aps,10pt,superscriptaddress,twocolumn,floatfix]{revtex4-1}
\usepackage{graphicx,color}
\usepackage[nice]{nicefrac}							
\usepackage{amsmath,amssymb,bm}
\usepackage{amsmath}
\usepackage{braket}
\usepackage{scrextend}

\usepackage[plainpages=false,pdfpagelabels,colorlinks=true,linkcolor=red,urlcolor=blue,citecolor=blue,pdftitle={},pdfauthor={},pdfdisplaydoctitle=true,pdfduplex=DuplexFlipLongEdge]{hyperref}

\definecolor{darkred}{rgb}{0.90,0.2,0.2}
\definecolor{darkgreen}{rgb}{0,0.60,.2}
\definecolor{darkblue}{rgb}{0.1,0.3,1}
\definecolor{grey}{cmyk}{0,0,0,0.25}
\definecolor{orange}{cmyk}{0,0.6,0.8,0}

\begin{document}

\title{Eigenstate Entanglement Entropy in Random Quadratic Hamiltonians}

\author{Patrycja  \L yd\.{z}ba}
\affiliation{Department of Theoretical Physics, J. Stefan Institute, SI-1000 Ljubljana, Slovenia}
\affiliation{Department of Theoretical Physics, Wroclaw University of Science and Technology, 50-370 Wroc{\l}aw, Poland}
\author{Marcos Rigol}
\affiliation{Department of Physics, The Pennsylvania State University, University Park, Pennsylvania 16802, USA}
\author{Lev Vidmar}
\affiliation{Department of Theoretical Physics, J. Stefan Institute, SI-1000 Ljubljana, Slovenia}
\affiliation{Department of Physics, Faculty of Mathematics and Physics, University of Ljubljana, SI-1000 Ljubljana, Slovenia}


\begin{abstract}
The eigenstate entanglement entropy is a powerful tool to distinguish integrable from generic quantum-chaotic models. In integrable models, the average eigenstate entanglement entropy (over all Hamiltonian eigenstates) has a volume-law coefficient that generally depends on the subsystem fraction. In contrast, it is maximal (subsystem fraction independent) in quantum-chaotic models. Using random matrix theory for quadratic Hamiltonians, we obtain a closed-form expression for the average eigenstate entanglement entropy as a function of the subsystem fraction. We test it against numerical results for the quadratic Sachdev-Ye-Kitaev model and show that it describes the results for the power-law random banded matrix model (in the delocalized regime). We show that localization in quasimomentum space produces (small) deviations from our analytic predictions.
\end{abstract}
\maketitle

{\it Introduction.}
Entanglement, a genuine property of the quantum world, provides unique ways of characterizing quantum many-body systems~\cite{amico_fazio_08, peschel_eisler_09, calabrese_cardy_09, eisert_cramer_10}. Studies of entanglement indicators have contributed novel insights into properties of ground states~\cite{audenaert_eisert_02, vidal_latorre_03}, quantum phase transitions~\cite{osterloh_amico_2002, osborne_nielsen_02}, information scrambling in nonequilibrium quantum dynamics~\cite{Kaufman2016, alba_calabrese_17}, and highly excited Hamiltonian eigenstates that exhibit eigenstate thermalization~\cite{deutsch_91, srednicki_94, rigol_dunjko_08} (see Refs.~\cite{eisert_friesdorf_15, dalessio_kafri_16, deutsch_18, mori_ikeda_18} for reviews). Models that display eigenstate thermalization and random matrix statistics in their spectrum are usually referred to as quantum-chaotic~\cite{dalessio_kafri_16}. Typical eigenstates of those models have a maximal von-Neumann entanglement entropy~\cite{beugeling_andreanov_15, vidmar_rigol_17, garrison_grover_18}.

Let $|m\rangle$ be an eigenket of a lattice Hamiltonian with two states per site (e.g., spinless-fermion or spin-1/2 Hamiltonians, our focus here) in one dimension. To compute the von-Neumann entanglement entropy (in short, the entanglement entropy) of $|m\rangle$, we bipartition the lattice with $L$ sites into a subsystem A with $L_{\rm A}$ contiguous sites and the environment B with $L-L_{\rm A}$ sites, and trace out the environment sites to obtain the reduced density matrix of subsystem A, $\hat \rho_{\rm A}^{(m)} = {\rm Tr}_{\rm B}\{|m\rangle \langle m |\}$. The entanglement entropy is $S_m = - {\rm Tr}\{\hat \rho_{\rm A}^{(m)} \ln \hat\rho_{\rm A}^{(m)}\}$.

For highly excited eigenstates of quantum-chaotic lattice Hamiltonians, the leading term in $S_m$ has been found to be proportional to $L_A$ (for $1 \ll L_{\rm A} \leq L/2$) and consistent with the thermodynamic entropy at the corresponding energy~\cite{deutsch_10, santos_polkovnikov_12, deutsch13, beugeling_andreanov_15, yang_chamon_15, vidmar_rigol_17, dymarsky_lashkari_18, garrison_grover_18, nakagawa_watanabe_18, lu_grover_19, huang_19, murthy_srednicki_19, wilming_goihl_19, miao_barthel_19, morampudi_chandran_20, faiez_sefranek_20_a, faiez_sefranek_20_b, bhakuni_sharma_2020, kaneko_iyoda_20}. Since the overwhelming majority of energy eigenstates in such systems is at ``infinite temperature''~\cite{dalessio_kafri_16}, this means that typical eigenstates have
\begin{equation} \label{def_S_qchaotic}
 S_m(L_A) \simeq \ln {\cal D}_{\rm A} = L_{\rm A} \ln 2 \,,
\end{equation}
where ${\cal D}_{\rm A} = 2^{L_{\rm A}}$ is the subsystem's Hilbert space dimension. $S_m(L_A)$ in Eq.~\eqref{def_S_qchaotic} matches the average~\cite{page_93}, as well as the typical~\cite{bianci_dona_19}, entanglement entropy of random pure states. The presence of conserved quantities, such as the particle number, only modifies subleading terms~\cite{vidmar_rigol_17}.

In sharp contrast, translationally invariant quadratic fermionic models (or models mappable to them) were proved to exhibit a qualitatively different behavior of the average and typical entanglement entropy of their many-body eigenstates~\cite{vidmar_hackl_17, vidmar_hackl_18, hackl_vidmar_19}. While the leading term in the average is still proportional to $L_A$ (for $1 \ll L_{\rm A} \leq L/2$), its magnitude depends on the subsystem fraction $f = L_{\rm A}/L$, and is smaller than the maximal value for $f>0$. Qualitatively similar results were obtained numerically for free fermions in a superlattice in two dimensions~\cite{storms_singh_14}, and for the translationally invariant interacting integrable spin-1/2 XXZ chain~\cite{leblond_mallayya_19}. The results in the latter model were very close (potentially identical in the thermodynamic limit) to the ones in Refs.~\cite{vidmar_hackl_17, vidmar_hackl_18, hackl_vidmar_19}.

Integrable models are characterized by extensive numbers of nontrivial conserved quantities~\cite{grabowski_mathieu_95, mierzejewski_prelovsek_15, ilievski_medenjak_2016}, and display distinct properties such as absence of thermalization when taken far from equilibrium~\cite{rigol_dunjko_07, ilievski15, essler_fagotti_2016, vidmar16}. This is attributed to the fact that their eigenstates do not exhibit eigenstate thermalization~\cite{rigol_dunjko_08, rigol_09, cassidy_clark_11, vidmar16, yoshizawa2018numerical, leblond_mallayya_19, mierzejewski_vidmar_20}. Lack of thermalization close to integrable points is robust enough as to be accessible in experiments with ultracold atoms~\cite{kinoshita_wenger_06, langen15a, langen_gasenzer_2016, tang_kao_18}, in which distributions of conserved quantities (rapidities) were recently measured~\cite{wilson_malvania_20}. It is then important to develop theoretical tools to distinguish integrable from quantum-chaotic models. The results in Refs.~\cite{vidmar_hackl_17, vidmar_hackl_18, hackl_vidmar_19, storms_singh_14, leblond_mallayya_19} show that the eigenstate entanglement entropy is one of such tools. In contrast to traditionally used spectral properties, it does not require finding and removing all symmetries of the model~\cite{dalessio_kafri_16}. Another recently used eigenstate-based tool, the AGP norm, involves studying the response of energy eigenstates to perturbations~\cite{Pandey:2020}.

For the eigenstate entanglement entropy, a stepping stone missing for integrable models that is available for quantum-chaotic ones is a closed-form expression for the average entanglement entropy, like the one in Ref.~\cite{page_93}, which could serve as a reference point to compare to results obtained for specific Hamiltonians. We provide such a stepping stone in this work by computing the average entanglement entropy of random quadratic Hamiltonians. We identify properties of the models to which we expect it to apply.

The closed-form expression is, for $f\leq 1/2$~\cite{fgt12},
\begin{equation} \label{def_S}
\bar {\mathcal S}(L_A,f) = \left[1 - \frac{1 + f^{-1}(1-f) \ln(1-f)}{\ln 2} \right] L_{\rm A} \ln 2 \,,
\end{equation}
and we obtain it using random matrix theory (RMT) for quadratic Hamiltonians. We test Eq.~\eqref{def_S} against numerical results for the quadratic Sachdev-Ye-Kitaev (SYK2) model, and provide evidence that it describes the average entanglement entropy in the delocalized regime of the power-law random banded matrix (PLRBM) model. We also show that localization in quasimomentum space (e.g., because of translational invariance) results in (small) deviations from Eq.~\eqref{def_S}. This is in stark contrast to quantum-chaotic systems, in which translational invariance is known not to affect the leading term in the average entanglement entropy~\cite{vidmar_rigol_17}. 

{\it Derivation of Eq.~\eqref{def_S}.}
We consider quadratic fermionic Hamiltonians which, after diagonalization, can always be written as $\hat H = \sum_{q=1}^L \varepsilon_q \hat c_q^\dagger \hat c_q$, where $\varepsilon_q$ are the single-particle eigenenergies and $\{|q\rangle = \hat c_q^\dagger |\emptyset\rangle \, ; \, q=1,...,L \}$ are the single-particle energy eigenkets. Let the unitary transformation between the energy eigenkets and the position eigenkets $\{|i\rangle = \hat f_i^\dagger |\emptyset\rangle \, ; \, i=1,...,L \}$ be carried out by a matrix with elements $v_{iq}$, so that $\hat f_i = \sum_{q=1}^L v_{iq} \hat c_q$. 

The many-body eigenkets of $\hat H$ can be written as $\{|m\rangle = \prod_{\{ q_l \}_m} \hat c_{q_l}^\dagger |\emptyset \rangle\, ; \, m=1,...,2^L \}$, where $\{ q_l \}_m$ represents the $m$th set of occupied single-particle energy eigenkets. Introducing $\hat N_q = 2\hat c_q^\dagger \hat c_q - 1$, for which $\hat N_q |m\rangle = N_q^m|m\rangle$ with $N_q^m=1$ ($-1$) for occupied (empty) single-particle eigenkets, we can write the generalized one-body correlation matrix (in short, the correlation matrix) as
\begin{equation} \label{def_J}
\left({\cal J}_m\right)_{ij} = 2 \,\langle m | \hat f_i^\dagger \hat f^{}_j | m \rangle  - \delta_{ij} = \sum_{q=1}^{L} N^m_q \, v_{iq}^* v_{jq}^{}\,,
\end{equation}
where $i,j\leq L_A$, and we denote the eigenvalues as $\{ \lambda_j^{(m)}  ; \; j=1,...,L_{\rm A} \}$. In Eq.~\eqref{def_J}, we used the orthonormality condition: $\delta_{ij} = \sum_{q=1}^{L} v_{iq}^* v_{jq}^{}$. Further on, we shorten the notation ${\cal J}_m \to {\cal J}$ and $\lambda_j^{(m)} \to \lambda_j$.

The entanglement entropy of many-body eigenket $|m\rangle$ can then be computed as~\cite{vidal_latorre_03, peschel_03},
\begin{equation}\label{eq:seigenvalues}
 {\mathcal S}_m\! =\! -\! \sum_{j=1}^{L_{\rm A}}\! \left(\frac{1+\lambda_j}{2}\! \ln\! \left[\frac{1+\lambda_j}{2}\right]\! +\! \frac{1-\lambda_j}{2}\! \ln\! \left[\frac{1-\lambda_j}{2}\right]\right)\!,
\end{equation}
and the average (over all eigenstates) entanglement entropy is defined as $\bar {\mathcal S} \equiv 2^{-L} \sum_{m=1}^{2^L} {\mathcal S}_m$.

In order to make analytic progress in the evaluation of $\bar {\mathcal S}$, one can write
\begin{equation} \label{def_S_avr}
\bar {\mathcal S} = L_{\rm A} \ln 2 - \sum_{n=1}^{\infty} \frac{\overline{{\rm Tr}\left\{ {\cal J}^{2n}\right\}}}{2n\left(2n-1\right)} \,.
\end{equation}
This series was proved to converge in Ref.~\cite{vidmar_hackl_17}. The fact that only even powers of the correlation matrix ${\cal J}$ appear in the series enabled the computation of upper and lower bounds for translationally invariant systems in Refs.~\cite{vidmar_hackl_17, vidmar_hackl_18, hackl_vidmar_19}. In this work we use Eq.~\eqref{def_S_avr} to obtain a closed-form expression for $\bar {\mathcal S}$ in random quadratic Hamiltonians.

Our central assumption is a random matrix theory (RMT) assumption for the single-particle energy eigenkets. We assume that $v_{iq} = u_{iq}/\sqrt{L}$, where $u_{iq}$ is a normally distributed complex variable with zero mean and unit variance. This is equivalent to assuming that the quadratic Hamiltonians are represented by random matrices drawn from the Gaussian unitary ensemble (GUE). We note that our (leading-term) results do not change if we assume $u_{iq}$ to be a normally distributed real variable with the same mean and variance or, equivalently, the Hamiltonians to be represented by matrices drawn from the Gaussian orthogonal ensemble (GOE)~\cite{suppmat}.

Let us use our assumption to evaluate the first trace ($n=1$) in the series in Eq.~\eqref{def_S_avr}. Using Eq.~(\ref{def_J}), we get
\begin{eqnarray}
\overline{{\rm Tr} \{ {\cal J}^{2} \} } 
& = & \frac{1}{L^2} \sum_{i,j=1}^{L_{\rm A}} \sum_{q_1,q_2 = 1}^{L}\overline{N^m_{q_1} N^m_{q_2}}\, u_{iq_1}^* u_{jq_1} u_{jq_2}^* u_{iq_2} \nonumber \\
& = & \frac{1}{L^2} \sum_{i,j=1}^{L_{\rm A}} \sum_{q=1}^{L} |u_{iq}|^2 |u_{jq}|^2 \, ,
\label{def_Tr_m1}
\end{eqnarray}
where the average over all $m$ is $\overline{N^m_{q_1} N^m_{q_2}}= \delta_{q_1,q_2}$. Then, the RMT assumption implies that $\sum_{a=1}^{L_{\rm A}} |u_{aq}|^2 = L_{\rm A}$ for $a=i,j$, which yields
\begin{equation} \label{def_m1}
\overline{{\rm Tr} \{ {\cal J}^{2} \}} = \frac{L_{\rm A}^2}{L} = L_{\rm A} f \,.
\end{equation}
Remarkably, this is the universal result one gets for translationally invariant systems~\cite{vidmar_hackl_17}.

We have also computed the averages of traces for powers $n=2,\,3$, and 4 (see Ref.~\cite{suppmat} for details), 
\begin{eqnarray}
\overline{ {\rm Tr} \{ {\cal J}^{4}\} } & = & L_{\rm A} \left( 2f^2-f^3 \right) \,, \label{def_m2} \\
\overline{ {\rm Tr} \{ {\cal J}^{6}\} } & = & L_{\rm A} \left( 5f^3-6f^4+2f^5 \right) \,, \label{def_m3} \\
\overline{ {\rm Tr} \{ {\cal J}^{8}\} } & = & L_{\rm A} \left[ 14f^4 + {\cal O}(f^5) \right] \,. \label{def_m4}
\end{eqnarray}
Plugging these results in Eq.~\eqref{def_S_avr}, we get
\begin{equation} \label{def_S_m3}
\bar {\mathcal S} = L_{\rm A} \ln 2 - L_{\rm A} \left[\frac{f}{2}+\frac{f^2}{6}+\frac{f^3}{12}+\frac{f^4}{20} + O(f^5)\right] \, .
\end{equation}
Equation~(\ref{def_S_m3}) is an exact expansion in $f$ up to ${\cal O}(f^5)$, since
$\overline{ {\rm Tr} \{ {\cal J}^{2n} \} } / L_{\rm A}$ is in general a polynomial that, in the thermodynamic limit, contains only powers from $f^n$ up to $f^{2n-1}$~\cite{vidmar_hackl_17}. The computation of the traces in Eq.~\eqref{def_S_avr} becomes increasingly tedious as $n$ increases~\cite{suppmat}. From the terms in Eq.~(\ref{def_S_m3}), we conjecture the series to be
\begin{equation}\label{eq:conjecture}
\bar{\mathcal S} = L_{\rm A} \ln 2 - L_{\rm A} \sum_{n=1}^{\infty} \frac{f^n}{n \left(n+1\right)} \,,
\end{equation}
which yields Eq.~(\ref{def_S}), our main result. A first test of the correctness of our conjecture is provided by $\bar{\mathcal S}(L_A,1/2)=(2\ln2-1)L_A$, as found in Ref.~\cite{SYK_Liu_2017}. We also note that Eq.~\eqref{def_S}, and its $f>1/2$ counterpart~\cite{fgt12}, share properties of Page's result~\cite{page_93}: $\bar{S}_P = L_{\rm A} \ln 2 - (1/2) {\cal D}_{\rm A}^2/2^L$, where ${\cal D}_{\rm A} = 2^{L_{\rm A}}$ if $L_{\rm A} \leq L/2$ and ${\cal D}_{\rm A} = 2^{L-L_{\rm A}}$ otherwise. When comparing the derivatives of $\bar {\cal S}$ and $\bar{S}_P$ at $f = 1/2$, for $f$ approaching 1/2 from below and above, the first derivative and all even derivatives of both expressions agree, while odd derivatives higher than the first do not. It is conceivable that the same way that Page's result describes the average over all pure states, our result~(\ref{def_S}) describes the average over all fermionic Gaussian states (see, e.g., Ref.~\cite{fukuda_koenig_19} for recent results on bosons). 

\begin{figure}[!tb]
\includegraphics[width=0.98\columnwidth]{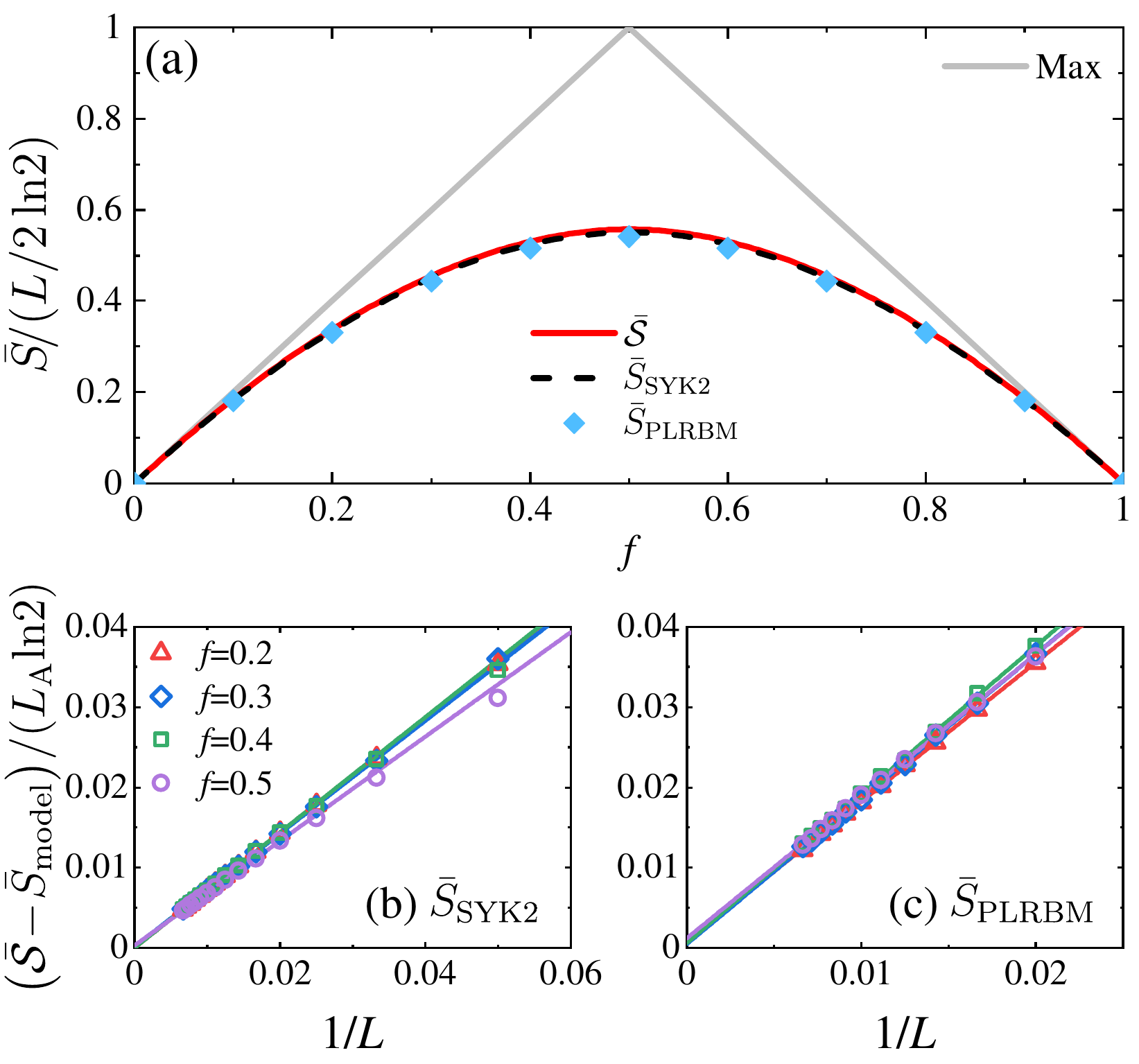}
\vspace{-0.2cm}
 \caption{(a) Average entanglement entropy $\bar S/(L/2 \ln2)$ as a function of the subsystem fraction $f$. Solid lines show the maximal value for random pure states~\eqref{def_S_qchaotic} and the RMT prediction $\bar{\mathcal S}$ for quadratic models~\eqref{def_S}, while the dashed line and symbols correspond to numerical results for the SYK2 and the PLRBM model, respectively, for $L=100$ lattice sites. The numerical results were obtained by averaging over $5\times 10^5$ many-body eigenstates for each of 500 Hamiltonian realizations. (b) and (c) Finite-size scaling of $(\bar{\mathcal S} - \bar{S}_{\rm model})/(L_{\rm A} \ln 2)$, where $\bar{S}_{\rm model}$ are the numerical results obtained for the SYK2 and the PLRBM model, respectively. Lines are fits of the results (for $L \geq 50$) to the function $a_0 + a_1/L$, with free parameters $a_0$ and $a_1$. We get $|a_0| \lesssim 10^{-3}$ in all cases.}
\label{fig1}
\end{figure}

{\it SYK2 and PLRBM models.}
In position space, our quadratic Hamiltonians (in chains with $L$ sites) read
\begin{equation}\label{eq:hamrealspace}
\hat{H} = \sum_{ij}^{L} A_{ij} \hat{f}_i^\dagger\hat{f}_{j}^{} \,,
\end{equation}
where $A_{ij} = A_{ji}^*$. In order to test the correctness of our closed-form expression for the eigenstate entanglement entropy average~\eqref{def_S}, we construct matrices ${\bf A}$ drawn from the GUE. Namely, matrices whose elements $A_{i\neq j}$ are i.i.d.~complex numbers whose real and imaginary part are normally distributed with zero mean and variance $1/L$, while the elements $A_{ii}$ are i.i.d.~normally distributed real numbers with zero mean and variance $2/L$. For such matrices ${\bf A}$, Eq.~\eqref{eq:hamrealspace} is known as the Dirac fermion version of the SYK2 model (in short, the SYK2 model). Due to particle-hole symmetry, the entanglement entropy of the SYK2 model does not change if one replaces the Dirac fermions with Majorana fermions~\cite{SYK_Liu_2017}. In spite of large interest in entanglement properties of typical eigenstates of the SYK2 model~\cite{SYK_Liu_2017, zhang_liu_20}, a closed-form expression for the subsystem fraction dependence has remained elusive. 

In Fig.~\ref{fig1}(a), we show that the numerical evaluation of the average eigenstate entanglement entropy of the SYK2 model, for a chain with $L=100$ sites, follows very closely the analytical prediction~\eqref{def_S}. The small differences between the two are due to finite-size effects. In Fig.~\ref{fig1}(b), finite-size scaling analyses for different values of $f$ reveal that the numerical results converge to the analytic predictions with increasing system size.

We also studied the average eigenstate entanglement entropy in the PLRBM model, which is a model that exhibits a delocalization-localization transition in one dimension~\cite{mirlin_fyodorov_96}. In that model, the matrix elements $A_{ij}$ are taken to be real numbers normally distributed with zero mean and an algebraically decaying variance ${\rm Var}(A_{ij}) = 1/[1+\left(|i-j|/\beta\right)^{2\alpha}]$. We use open boundaries and focus on the delocalized regime by setting $\alpha=0.2$ and $\beta=0.1$~\cite{mirlin_fyodorov_96}. The results for $L=100$, also shown in Fig.~\ref{fig1}(a), closely follow our analytic prediction. Once again the deviations are due to finite-size effects, which are stronger than for the SYK2 model. In Fig.~\ref{fig1}(c), finite-size scaling analyses for different values of $f$ reveal that the numerical results for the PLRBM model converge to the predictions of Eq.~\eqref{def_S} with increasing system size.

\begin{figure}[!t]
\includegraphics[width=0.98\columnwidth]{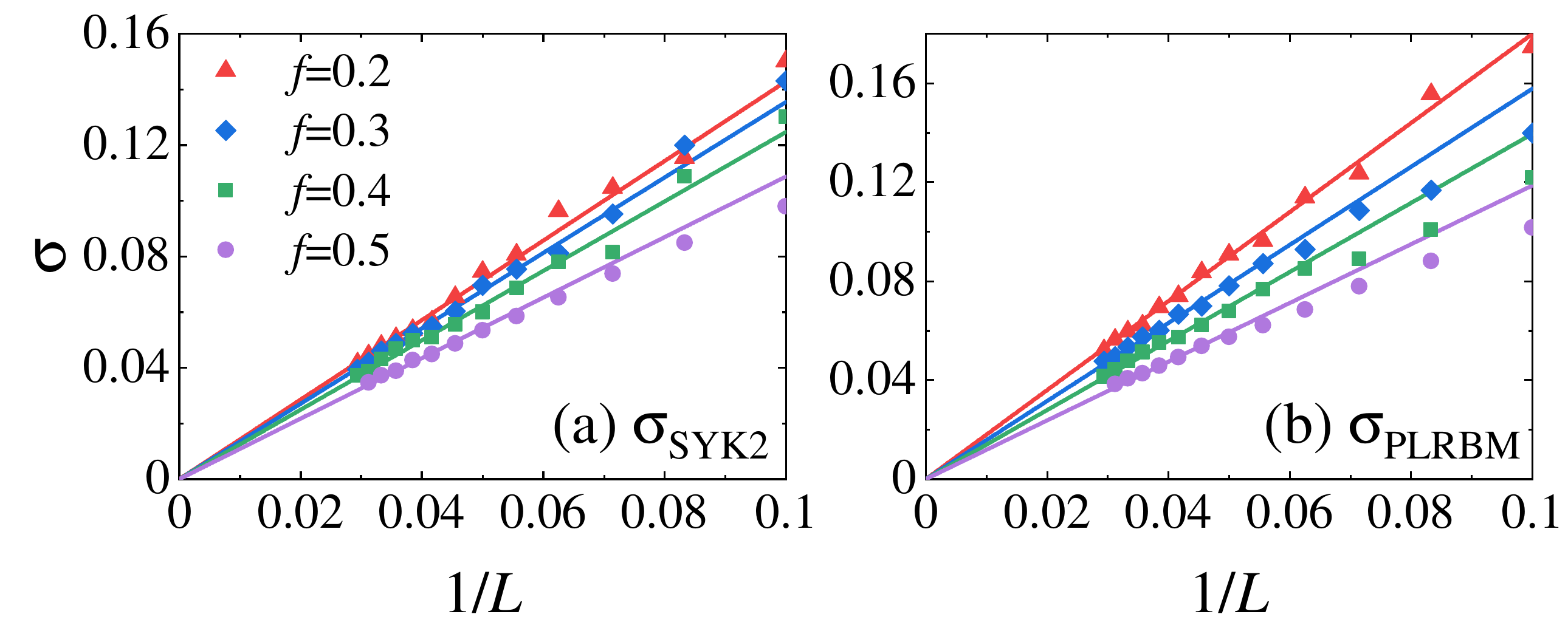}
\vspace{-0.2cm}
 \caption{Variances $\sigma$ versus $1/L$ for (a) the SYK2 model, and (b) the PLRBM model, at different subsystem fractions $f$ with $L_{\rm A}$  being the closest integer to $fL$. For a given Hamiltonian realization, we define $\sigma_\mu^2 = (\overline{S^2} - \overline{S}^2)/(L_{\rm A} \ln 2)$, where the average is carried out over all many-body eigenstates. We then obtain $\sigma$ by averaging $\sigma_\mu$ over $M$ Hamiltonian realizations ($M=50$ for $L>16$, $M=100$ for $L\leq16$). Lines are fits to a single-parameter function $a_1/L$ for $L \geq 20$.}
\label{fig2}
\end{figure}

For both the SYK2 and PLRBM models, we find that the variances $\sigma$ of the distributions of eigenstate entanglement entropies for $f>0$ vanish in the thermodynamic limit, see Fig.~\ref{fig2}. This means that the typical eigenstate entanglement entropy is the same as the average. Hence both terms can be used interchangeably. For both models, we find the scaling $\sigma \propto 1/L$ for large $L$. It is interesting to note that an identical scaling was found in Ref.~\cite{vidmar_hackl_17} for translationally invariant free fermions in the limit $f\to 0$, for which it was shown that the distribution of eigenvalues of the correlation matrices ${\cal J}$ is described by a Toeplitz Gaussian ensemble.

{\it Localization in quasimomentum space.}
Many-body eigenstates of models that are localized in position space have an area law entanglement entropy. Hence, the average and typical eigenstate entanglement entropy in those models is not described by Eq.~\eqref{def_S}. In what follows we discuss what happens for quadratic models that exhibit localization in quasimomentum space.

First, we consider spinless fermions with nearest neighbor hoppings in a homogeneous lattice with either periodic (PBCs) or open (OBCs) boundary conditions,
\begin{equation}
\label{eq4}
\hat{H}_{\rm PBC/OBC}\! =\! -\sum_{i=1}^{L-1} \left( \hat{f}_{i}^\dagger \hat{f}_{i+1}\! +\! \hat{f}_{i+1}^\dagger \hat{f}_{i} \right)\! -\! \eta \left( \hat{f}_{L}^\dagger \hat{f}_{1}\! +\! \hat{f}_{1}^\dagger \hat{f}_{L} \right)\! ,
\end{equation}
where $\eta=1$ for $\hat H_{\rm PBC}$ and $\eta=0$ for $\hat H_{\rm OBC}$. Previous studies focused on the entanglement entropy of excited eigenstates in these and related models~\cite{alba_fagotti_09, storms_singh_14, moelter_barthel_14, lai_yang_15, nandy_sen_16, vidmar_hackl_17, riddell_muller_18, vidmar_hackl_18, hackl_vidmar_19, jafarizadeh_rajabpour_19}. 

\begin{figure}[t]
\includegraphics[width=0.98\columnwidth]{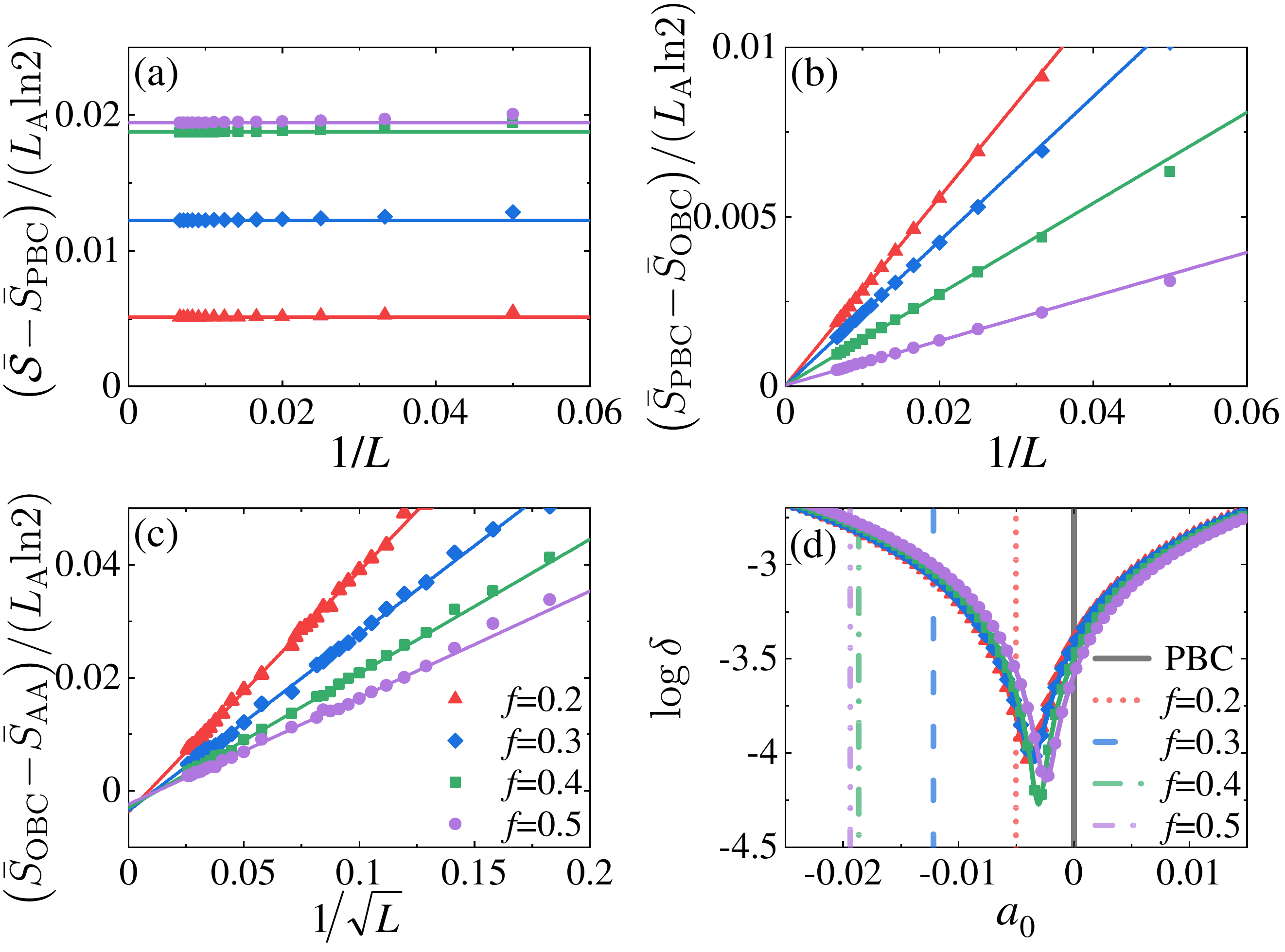}
\vspace{-0.2cm}
 \caption{Finite-size scalings at different subsystem fractions $f$. (a) $(\bar {\mathcal S}-\bar S_{\rm PBC})/(L_{\rm A} \ln 2)$ vs $1/L$. Lines are fits to a constant. (b) $(\bar S_{\rm PBC}-\bar S_{\rm OBC})/(L_{\rm A} \ln 2)$ vs $1/L$. Lines are fits to a two-parameter function $a_0 + a_1/L$. We get $|a_0| < 4 \times 10^{-5}$ in all cases. The numerical results in (a) and (b) were obtained averaging over at least $10^6$ many-body eigenstates. (c) $(\bar S_{\rm OBC}-\bar S_{\rm AA})/(L_{\rm A} \ln 2)$ vs $1/\sqrt{L}$. The results for $\bar S_{\rm AA}$ are averaged over $10^4 - 10^6$ many-body eigenstates for a single realization of $\hat H_{\rm AA}$, followed by 100-500 realizations with random parameters $W \in [0.55,0.65]$ and $\phi \in [0,2\pi)$. Lines are fits to a two-parameter function $g(L)=a_0 + a_1/\sqrt{L}$. In panels (a)--(c), we use results for $L\geq 70$ in the fits. (d) Quality of the fits in panel (c), defined as $\delta^2 = \sum_j [\delta s(L_j) - g(L_j)]^2/N $, where $\delta s = (\bar S_{\rm OBC}-\bar S_{\rm AA})/(L_{\rm A} \ln 2)$ and $N$ is the number of fitted points. We fix the intercept $a_0$ in the function $g(L)$, and plot $\delta$ vs $a_0$. Vertical dashed lines depict $(\bar S_{\rm PBC} - \bar {\mathcal S})/(L_{\rm A} \ln 2)$.}
\label{fig3}
\end{figure}

The volume-law coefficient of the average entanglement entropy $\bar S_{\rm PBC}/(L_{\rm A} \ln 2)$ of $\hat H_{\rm PBC}$ can be calculated numerically very accurately in the thermodynamic limit (the finite-size effects are exponentially small in $L$)~\cite{hackl_vidmar_19}. Figure~\ref{fig3}(a) depicts a finite-size scaling analysis of the difference between the latter and the volume-law coefficient in Eq.~\eqref{def_S}, $\bar{\mathcal S}/(L_{\rm A} \ln 2)$. This analysis shows that they are (slightly) different. Such a difference highlights that, contrary to quantum-chaotic systems in which localization in quasimomentum space does not affect the volume-law coefficient, in quadratic systems it does~\footnote{In Ref.~\cite{suppmat} we confirm that, as expected, the eigenkets of the SYK2 and the PLRBM models are delocalized in both position and quasimomentum space.}. This is consistent with the analytic observation that the Maclaurin series of $\bar {\mathcal S}$ and $\bar S_{\rm PBC}$ (cf.~Eq.~(57) in Ref.~\cite{hackl_vidmar_19}) differ starting with the term proportional to $f^2$. Using OBCs one can relax the condition of the single-particle energy eigenstates being quasimomentum eigenstates, while still keeping them localized in quasimomentum space. The results in Fig.~\ref{fig3}(b) suggest that $\bar S_{\rm OBC}/(L_{\rm A} \ln 2) \to \bar S_{\rm PBC}/(L_{\rm A} \ln 2)$ in the thermodynamic limit, namely, that OBCs do not affect the volume law.

The second example is the quasiperiodic Aubry-Andr{\'e} model~\cite{Aubry_1979}
\begin{equation} \label{def_Haa}
\hat{H}_{\rm AA} = \hat H_{\rm OBC} +  W \sum_{i=1}^{L} \cos\left(2\pi\sigma i + \phi\right)\hat{f}_{i}^\dagger\hat{f}_{i}^{} \,,
\end{equation}
whose eigenstate entanglement properties have also been studied in the past~\cite{li_pixley_16, Roy_2019, modak_nag_20}. We set $\sigma=\left(\sqrt{5}-1\right)/2$ to ensure incommensurability of the potential with the lattice periodicity, and $\phi$ is a constant. In the regime we are interested in, namely $W<2$, the single-particle energy eigenstates are delocalized in position space and localized in quasimomentum space~\cite{Aubry_1979, he_santos_13}.

We find that the subleading corrections to the volume-law coefficient of the average entanglement entropy $\bar S_{\rm AA}/(L_{\rm A} \ln 2)$ are much larger than in the homogeneous models. To subtract the effect of OBCs, we compute $(\bar S_{\rm OBC}-\bar S_{\rm AA})/(L_{\rm A} \ln 2)$ and plot it vs $1/\sqrt{L}$ in Fig.~\ref{fig3}(c). The results there suggest that the first subleading term in $\bar S_{\rm AA}$ (for $f>0$) is $\propto \sqrt{L_{\rm A}}$. We estimate the volume-law coefficient by finding the minimum of the quality $\delta$ of the linear fits in Fig.~\ref{fig3}(c). The results in Fig.~\ref{fig3}(d) strongly suggest that $\bar S_{\rm AA}$ is not described by $\bar {\mathcal S}$ in Eq.~\eqref{def_S} (see the differences between the positions of minima of $\delta$ and the vertical dashed lines). The leading volume-law term in $\bar S_{\rm AA}$ is very close (possibly the same) as for  $\bar S_{\rm PBC/OBC}$.

{\it Summary.}
We report a closed-form expression for the average eigenstate entanglement entropy $\bar {\mathcal S}$ of random quadratic Hamiltonians [Eq.~\eqref{def_S}], which can be seen as the RMT free-fermion counterpart to Page's result~\cite{page_93}. We tested it against numerical results for the SYK2 model (and an analytical result for $f=1/2$~\cite{SYK_Liu_2017}). We showed that $\bar {\mathcal S}$ describes the average and typical entanglement entropy in the PLRBM model (in the delocalized phase), and that localization in quasimomentum space leads to (small) deviations from $\bar {\mathcal S}$. As a result, we expect $\bar {\mathcal S}$ to describe the average eigenstate entanglement entropy of quadratic models whose single-particle eigenstates are not only delocalized but also sufficiently random in the position basis. We expect that correlations in the coefficients of delocalized single-particle wave functions in the position basis, such as those generated by localization in quasimomentum space, will lead to (small) deviations from $\bar {\mathcal S}$. Our results provide a stepping stone for studies of the average and typical entanglement entropy of eigenstates of integrable models, whose structure has been recently unveiled as being much richer than the one of quantum-chaotic models~\cite{leblond_mallayya_19, Pandey:2020, Brenes:2020}. 

\acknowledgements
We thank L. Hackl for comments and questions about our results. This work was supported by the Slovenian Research Agency (ARRS), Research core fundings Grants No.~P1-0044 and No.~J1-1696 (P.\L.~and L.V.) and by the National Science Foundation under Grant No.~PHY-2012145 (M.R.).

\bibliographystyle{biblev1}
\bibliography{references}

\end{document}